\documentclass[aps,twocolumn,preprintnumbers,showpacs,nofootinbib]{revtex4-1}

\usepackage{color} 

\definecolor{green}{rgb}{0,.5,0}

\definecolor{red}{rgb}{1,0,0}

\usepackage{graphicx}
\usepackage[subfigure]{graphfig}
\usepackage{epsfig}
\usepackage{dcolumn}
\usepackage{amsmath}

\newcommand{\xg}{\langle x \rangle_g}
\newcommand{\ocT}{{\overline{\mathcal T}}}
\newcommand{\ZRI}{Z^\text{RI}}
\newcommand{\MSbar}{{\overline{\text{MS}}}}

\DeclareMathOperator{\Tr}{Tr}

\begin{document}

\preprint{MSUHEP-18-005}

\title{Nonperturbatively-renormalized glue momentum fraction at the physical pion mass from Lattice QCD}

\author{{Yi-Bo Yang$^{1\dagger}$, Ming Gong$^{2\dagger}$, Jian Liang$^{3\dagger}$, Huey-Wen Lin$^{1\dagger}$, Keh-Fei Liu$^{3\dagger}$, Dimitra Pefkou$^4$ and Phiala Shanahan$^{4,5}$}
\vspace*{-0.5cm}
\begin{center}
\large{
\vspace*{0.4cm}
\vspace*{0.4cm}
($^\dagger$$\chi$QCD Collaboration)
}
\end{center}
}
\affiliation{
$^{1}$\mbox{Department of Physics and Astronomy, Michigan State University, East Lansing, MI 48824, USA}\\
$^{2}$\mbox{Institute of High Energy Physics, Chinese Academy of Sciences, Beijing 100049, China}\\
$^{3}$\mbox{Department of Physics and Astronomy, University of Kentucky, Lexington, KY 40506, USA}\\
$^{4}$\mbox{Department of Physics, College of William and Mary, Williamsburg, VA 23187-8795, USA}\\
$^{5}$\mbox{Thomas Jefferson National Accelerator Facility, Newport News, VA 23606, USA}\\
}

\begin{abstract}
We present the first nonperturbatively-renormalized determination of the glue momentum fraction $\langle x \rangle_g$ in the nucleon, based on lattice-QCD simulations  at the physical pion mass using the cluster-decomposition error reduction (CDER) technique. We provide the first practical strategy to renormalize the gauge energy-momentum tensor (EMT) nonperturbatively in the RI/MOM scheme, and convert the results to the $\MSbar$ scheme with 1-loop matching. The simulation results show that the CDER technique can reduce the statistical uncertainty of its renormalization constant by a factor of ${\cal O}$(300) in calculations using a typical state-of-the-art lattice volume, and the nonperturbatively-renormalized $\xg$ is shown to be independent of the lattice definitions of the gauge EMT up to discretization errors. We determine the renormalized $\xg^{\overline{\textrm{MS}}}(2\textrm{ GeV})$ to be 0.47(4)(11) at the physical pion mass, which is consistent with the experimentally-determined value. 
\end{abstract}

\maketitle

 \section{Introduction}
 
  A longstanding problem raised by deep-inelastic scattering (DIS) and Drell-Yan experiments on the nucleon is that the gluons contribute almost as large a fraction of the nucleon momentum as the quarks~\cite{Gluck:1999xe,Dulat:2015mca}, contradicting the naive quark model. The momentum fractions of the quarks and glue equal the second moments of their respective parton distribution functions (PDFs) $f_p(x)$ ($p=u,\bar{u},d,\bar{d},s,...,g$):
\begin{equation}
\langle x \rangle_{p}=\int^1_0 dx\,x f_p(x),
\end{equation}
where the PDF can be determined from global fits of experimental results with certain assumptions about their functional forms. The recent CT14NNLO global PDF fit~\cite{Dulat:2015mca} yields $\xg^{\MSbar}(2\text{ GeV})=0.42(2)$, and the value at the TeV scale will be around 0.5 which is irrespective of its value at lower scales. Beside the importance in understanding the nucleon momentum,  the value of $\xg$ is also an important input to obtain the glue contributions to the nucleon mass and spin~\cite{Ji:1994av,Deka:2013zha}, so calculating it from a first-principle lattice-QCD simulation is of fundamental interest, in addition to providing an independent input and check of the experimental PDF determinations.

Lattice calculations of $\xg$ in the nucleon~\cite{Horsley:2012pz,Deka:2013zha,Alexandrou:2016ekb,Alexandrou:2017oeh} have been significantly refined in the last 10 years.
However,  values of $\xg^{\MSbar}(2\text{ GeV})$ vary widely, two quenched calculations found 0.43(9) and 0.33(6)~\cite{Horsley:2012pz,Deka:2013zha}, and recent dynamical $N_f=2$ calculation obtained 0.267(22)(30)~\cite{Alexandrou:2016ekb,Alexandrou:2017oeh}. 

The recent quenched (Refs.~\cite{Horsley:2012pz,Deka:2013zha}) and dynamical (Refs.~\cite{Alexandrou:2016ekb,Alexandrou:2017oeh}) lattice calculation of $\xg$ used different lattice definitions of the gauge energy-momentum tensor (EMT) with the 1-loop renormalization based on the lattice perturbation theory (LPT). It is known that LPT is poorly convergent at 1-loop level without smearing of the gauge EMT~\cite{Corbo:1989yj,Capitani:1994qn}, and LPT calculations beyond 1-loop level are extremely difficult. Whether smearing of the gauge EMT can improve the convergence of LPT remains an open question, but it was found in Ref.~\cite{Yang:2016plb} that hypercubic (HYP) smearing~\cite{Hasenfratz:2001tw} of the glue operator can change the bare glue matrix element by a factor of $\sim$3. Nonperturbative renormalization (NPR) of $\xg$ is thus essential to check whether different lattice definitions of the gauge EMT and smearing can provide a consistent prediction of $\xg$.

In this work, we present the first NPR of the gauge EMT using the cluster-decomposition error reduction (CDER). We confirm that, nonperturbatively-renormalized $\xg$ is independent of the lattice definition of the gauge EMT and whether the HYP smearing is applied to it. 

The glue NPR technique that we introduce will be applicable for the quantities beyond the $\xg$. State-of-the-art calculations of the glue spin contribution to the proton spin~\cite{Yang:2016plb} and the glue transversity in hadrons~\cite{Detmold:2016gpy} have been presented recently, the renormalization of the glue operators in these calculations are determined at the 1-loop level or neglected entirely. Approaches that target the entire glue PDF instead of the moments, like large-momentum effective theory (LaMET)~\cite{Ji:2014gla} and the lattice cross-section approach~\cite{Ma:2017pxb}, have been explored recently. NPR will be also essential to obtain accurate predictions for those quantities.

{In the rest part of the paper, we will start from the simulation strategy of NPR in Sec.~\ref{sec:npr}. Then in Sec.~\ref{sec:tests}, this strategy is tested in several cases including the quenched, 2 flavor, and 2+1 flavor ones. Based on those tests, a prediction of the renormalized $\xg$ is provided in Sec.~\ref{sec:result}, with controllable systematic uncertainties from NPR. Our findings in this work are summarized in Sec.~\ref{sec:summary}, and the additional discussion on the cases with more than 1-step of HYP smearing is presented in the appendix.}

\section{NPR Simulation Strategy}\label{sec:npr}

 At tree level, the gauge EMT $\ocT_{g,\mu\nu}\equiv F_{\mu\rho}F_{\nu\rho}-\frac{1}{4}g_{\mu\nu}F^2$ includes 9 Lorentz structures,
\begin{align}\label{eq:g_tree}
\ocT^{(0)}_{g,\mu\nu} =
  \big(2p_{\mu}p_{\nu}g_{\rho\tau} - p_{\mu}p_{\rho}g_{\nu\tau}
       +p^2g_{\rho\mu}g_{\nu\tau} - p_{\tau}p_{\nu}g_{\rho\mu} \nonumber\\
       -p_{\nu}p_{\rho}g_{\mu\tau} + p^2g_{\rho\nu}g_{\mu\tau} -
       p_{\tau}p_{\mu}g_{\rho\nu} \nonumber\\
       +g_{\mu\nu}\left(p_{\tau}p_{\rho}-p^2g_{\tau\rho}\right)\big)
  A_{\rho}(p)A_{\tau}(-p) ,
\end{align}
where $\mu$ and $\nu$ denote the external Lorentz indices of the EMT, $\rho$ (or $\tau$) is the Lorentz index of the external gluon state $A_{\rho/\tau}$. 
As discussed in Ref.~\cite{Collins:1994ee}, $2p_{\mu}p_{\nu}g_{\rho\tau}$ is the only structure free of mixing with the unphysical terms of the gauge EMT (gauge dependent term and ghost term), and is thus the best choice to consider the renormalization of the gauge EMT without the mixing calculation with unphysical terms.

{While taking the physical condition $p_{\rho}=p_{\tau}=0,$ \mbox{$p^2=0$}~\cite{Collins:1994ee} in the Minkowski space will isolate this term, the on-shell condition
$p^2 =0$ is not satisfied on the lattice. One can however choose other conditions on the lattice to isolate this term. 
 More precisely, the RI/MOM renormalization constant  of the off-diagonal pieces of the gauge EMT at the renormalization scale $\mu_R^2=p^2$ can be defined using the following approach, which is analogous to that commonly used for the quark bilinear operators~\cite{Martinelli:1994ty}:

\begin{align}\label{eq:Z_off}
\begin{split}
Z^{-1}(\mu_R^2) ={}&
\left(\frac{N_c^2-1}{2}\ZRI_g(\mu_R^2)\right)^{-1} \\
{}\times{}& \left.
\frac{V\langle \ocT_{g,\mu\nu}\Tr[A_{\rho}(p) A_{\rho}(-p)]\rangle}
     {2p_{\mu}p_{\nu}\langle \Tr[A_{\rho}(p) A_{\rho}(-p)]\rangle^2}
\right|_{\tiny{\substack{p^2=\mu_R^2,\\\rho\neq\mu\neq\nu,\\p_\rho=0}}}
\end{split}\nonumber \\
{}=& \left.
\frac{p^2\langle \ocT_{g,\mu\nu}\Tr[A_{\rho}(p) A_{\rho}(-p)]\rangle}
{2p_{\mu}p_{\nu}\langle\Tr[A_{\rho}(p) A_{\rho}(-p)]\rangle}
\right|_{\tiny{\substack{p^2=\mu_R^2,\\\rho\neq\mu\neq\nu,\\p_\rho=0}}},
\end{align}
where $\ocT_{g,\mu\nu}$ the index $\rho$ is not summed and $V$ is the physical volume of the lattice. 
The final expression in the right-hand side of Eq. (3) does
not depend on the renormalization constant 
\begin{align}
\ZRI_g\frac{\langle \Tr[A_{\rho}(p) A_{\rho}(-p)]\rangle }{V}=\frac{N_c^2-1}{2} \frac{1}{p^2}.
\end{align}
 in the RI/MOM scheme, as it is cancelled by the inverse of the $\langle \Tr[A_{\rho}(p) A_{\rho}(-p)]\rangle$ in its definition.

The Landau gauge-fixed gluon field $A_{\rho}(p)$ used above is defined from the gauge links $U_{\mu}(x)$ as:
\begin{align}\label{eq:def}
A_{\rho}(p) = a^4\sum_{x} e^{ip\cdot(x+\frac{1}{2}\hat{\rho})}
  \left[\frac{U_\rho(x)-U^{\dagger}_\rho(x)}{2ig_0a}\right]_\text{traceless}.
\end{align}
Note that even though the operator $\ocT$ may be HYP smeared, no smearing will be applied to the gauge field $A_{\rho}(p)$, since the gauge action is not smeared and no reweighting is applied to the configurations. 
Similarly, the RI/MOM renormalization constants of the traceless diagonal pieces of the gauge EMT can be defined by:
\begin{multline}\label{eq:Z_tr}
Z_T^{-1}(\mu_R^2)= \\
\left.
\frac{p^2\langle (\ocT_{\mu\mu}-\ocT_{\nu\nu})\Tr[A_{\rho}(p) A_{\rho}(-p)]\rangle}
{2p^2_{\mu}\langle\Tr[A_{\rho}(p) A_{\rho}(-p)]\rangle}\right|_{\tiny{\substack{p^2=\mu_R^2,\\ \rho\neq\mu\neq\nu, \\p_\rho=0,\\p_\nu=0}}}.
\end{multline}

The bare lattice gauge EMT can be defined by the clover definition of the field tensor $F_{\mu\nu}$~\cite{Horsley:2012pz,Deka:2013zha},
\begin{align}
\ocT^{(a)}_{g,\mu\nu} &= 2a^4\sum_x\Tr\left[F_{\mu\rho}F_{\nu\rho}-\tfrac{1}{4}g_{\mu\nu}F^2\right](x),\nonumber\\
F_{\mu\nu}(x) &= \frac{i}{8a^2g} \left[\mathcal{P}_{[\mu,\nu]}+\mathcal{P}_{[\nu,-\mu]}+ \mathcal{P}_{[-\mu,-\nu]} + \mathcal{P}_{[-\nu,\mu]} \right](x), \label{eq:emt_clover}
\end{align}
where the plaquette $\mathcal{P}_{\mu,\nu}(x)= U_\mu(x)U_\nu(x+a\hat{\mu})U^{\dagger}_\mu(x+a\hat{\nu})U^{\dagger}_\nu(x)$ with $U_{-\nu}(x)=U^{\dagger}_{\nu}(x-a\hat{\nu})$ and $P_{[\mu,\nu]}\equiv P_{\mu,\nu}-P_{\nu,\mu}$. The bare traceless diagonal component $\ocT_{g,\mu\mu}$ also has a simpler definition (the plaquette definition)~\cite{Alexandrou:2016ekb,Alexandrou:2017oeh}:
\begin{equation}\label{eq:emt_plq}
 \ocT^{(b)}_{g,\mu\mu} = \frac{-4}{g^2}
\Big(\sum_{\nu\neq\mu,x}\Tr[\mathcal{P}_{\mu,\nu}(x)]-\frac{1}{4}\sum_{\rho\neq\nu,x}\Tr[\mathcal{P}_{\rho,\nu}(x)]\Big).
\end{equation}
Different definitions and choices of smearing on the links $U_{\mu}(x)$ in these definitions of $\ocT_g$ yield different bare hadron matrix elements, but the renormalized results should agree up to ${\cal O}(a^2)$ correction.

After the renormalization constant $Z^{-1}(\mu_R^2)$ is obtained perturbatively or nonperturbatively under the lattice regularization at $\mu^2_R=p^2$, the matching factor to convert the result to the $\MSbar$ scheme should be calculated using dimensional regularization. At the $\mu_R$ used in this work, the 1-loop corrections to match the $\MSbar$ scheme at 2~GeV are at a few percent level~\cite{Yang:2016xsb}. The mixing with the quark EMT is also small~\cite{Yang:2016xsb} and will be considered as a systematic uncertainty; more detailed discussions of the matching and mixing effects can also be found there.

\begin{figure}[ht]
\includegraphics[scale=0.7]{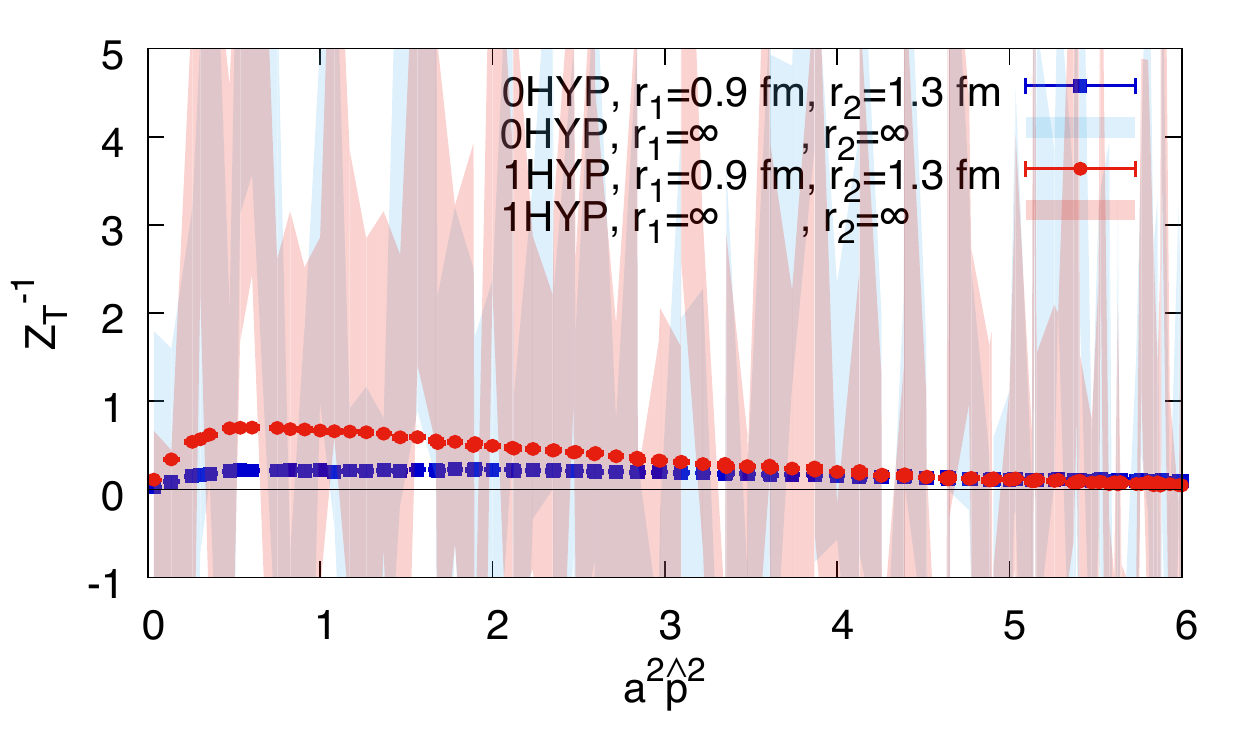}
\caption{The glue operator renormalization constants $Z^{-1}_T$ in $\MSbar$ at 2~GeV with and without CDER (i.e., cutoffs on the distance between the gauge fields/operator). Without CDER the errors are large and the signal cannot be resolved (bands in the background). The errors can be reduced by a factor of  $\sim$300 with $r_1=0.9$~fm, $r_2=1.3$~fm, shown by the red dots (blue boxes) for $Z^{-1}_T$ with (without) HYP smearing.}\label{fig:Z_op}
\end{figure}

Calculation of the correlation function 
\begin{align}
C_3(p)={}&\left\langle \ocT_{\mu\nu}\Tr[A_{\rho}(p) A_{\rho}(-p)]\right\rangle \nonumber\\
={}&\left\langle \int\!\! d^4x\, d^4y\, d^4z\, e^{ip (x-y)} \ocT_{\mu\nu}(z) \Tr[A_{\rho}(x) A_{\rho}(y)] \right\rangle\label{eq:def_ori}
\end{align}
is numerically challenging, even when the gluon propagator has been determined at better than the 1\% level.  Fig.~\ref{fig:Z_op}
illustrates this difficulty: the light-colored bands in the background show the direct calculations of $Z^{-1}_T(a^2 \hat{p}^2\equiv 4\sum_{\mu}\textrm{sin}^2\frac{ap_{\mu}}{2})$ (with the condition that two components of $p$ are zero and $\sum_{\mu} p_{\mu}^4/(\sum_{\mu} p_{\mu}^2)^2<$0.55) based on the
definition in Eq.~(\ref{eq:def_ori}), on 356 configurations of the 2+1-flavor RBC/UKQCD domain-wall fermion (DWF) Iwasaki gauge ensemble ``48I'' with lattice spacing $a=0.114$~fm, $m_{\pi}=140$~MeV and lattice volume $L^3\times V=48^3\times 96$ ($L=5.5$~fm)~\cite{Blum:2014tka}. The statistical uncertainties are very large and $Z^{-1}_T$ can not be resolved at any scale. 

However, we can apply the cluster-decomposition error reduction (CDER) technique to reduce the errors~\cite{Liu:2017man}.  The cluster-decomposition principle enunciates that correlerators falloff exponentially in the distance between operator insertions, and implies that integrating the correlator over this distance beyond the correlation length will only garner noise not signal. The CDER technique will cut off the volume integral beyond a characteristic length, and then one can gain a factor of $\sqrt{V}$ in the signal to noise ratio Ref.~\cite{Liu:2017man}.
Applying CDER to $C_3(p)$ in Eq.~(\ref{eq:def_ori}) introduces two cutoffs, $r_1$ between the glue operator and one of the gauge fields, and $r_2$ between the gauge fields in the gluon propagator, 
and then leads to the cutoff correlator:
\begin{align}
C_3^\text{CDER}(p) \equiv \Big\langle \int_{|r|<r_1}\!\!\!\!\!\!d^4r \int_{|r'|<r_2}\!\!\!\!\!\!d^4r' \int d^4x \nonumber\\
e^{ip\cdot r'} \ocT_{\mu\nu}(x+r) \Tr[A_{\rho}(x) A_{\rho}(x+r')] \Big\rangle. \label{eq:def_cdp}
\end{align}
For example, with cutoffs $r_1$=0.9~fm, $r_2$=1.3~fm, the statistical uncertainty can be reduced by a factor of approximately 300. 
This is close to the square root of $V^2$ over the product of 4-D spheres with radius $r_1$ and $r_2$,  $2V/(\pi^2r_1^2r_2^2)\simeq 263$. Using these parameters, a very clear signal can be resolved, shown as the red dots and blue boxes in Fig.~\ref{fig:Z_op}, for $Z^{-1}_T$ with and without HYP smearing, respectively. The values of $Z^{-1}_T$ differ by a factor of $\sim$3 for the calculations with or without the HYP smearing, at $a^2 \hat{p}^2\sim1$.

A naive cost estimate for the partial triple sum on the volume $V$ in Eq.~\ref{eq:def_cdp} is ${\cal O}(Vr_1^4r_2^4)$, but the practical cost can be reduced to ${\cal O}(V\log V)$ by applying the fast Fourier transform several times~\cite{Liu:2017man} {using the following strategy:

\begin{enumerate}
\item Construct $O^{r'}_{\mu\nu}(x)=\int_{|r'|<r_2} d^4r'\ocT_{\mu\nu}(x+r')$ by Fourier transforming  $\ocT_{\mu\nu}(x)$ and $f(x)=\theta(r_2-|x|)$, multiply the transformed functions together in momentum space, and then perform the anti-FT.

\item Calculate $B^{r'}_{\rho\mu\nu}(x)=A_{\rho}(x) \ocT_{\mu\nu}^{r'}(x)$.

\item Apply the cluster decomposition to $\int d^4x d^4y  e^{ip\cdot(x-y)} B_{\rho\mu\nu}^{r'}(x) A_{\rho}(y)$ \cite{Liu:2017man}: perform the FT for both $A$ and $B$, applying the anti-FT to $A(p)B(-p)$, apply the cut $g(x)=\theta(r_1-|x|)$ in coordinate space and then FT the product.
\end{enumerate}

The CDER with symmetric cutoffs 
\begin{align}
C_3(p)& \approx \Big\langle \int_{|r|<r_1}\!\!\!\!\!\!d^4r \int_{|r''|<r_3}\!\!\!\!\!\!d^4r'' \int d^4x \nonumber\\
&\,e^{ip\cdot (r+r'')} \ocT_{\mu\nu}(x) \Tr[A_{\rho}(x-r) A_{\rho}(x+r'')] \Big\rangle 
\end{align}
can also be efficient if a $V\log V$ implementation can be obtained.}

\section{Tests on CDER}\label{sec:tests}
{
Since the number of configurations in the 48I ensemble at $m_{\pi}$=140 MeV is limited, we turn to three ensembles with smaller volume and larger statistics to check the systematic uncertainties of the CDER approach.  To reduce statistical uncertainties then provide a stronger check, we will apply 1 step of HYP smearing on the gauge EMTs used in this section. 

\begin{figure}[ht]
  \includegraphics[scale=0.7]{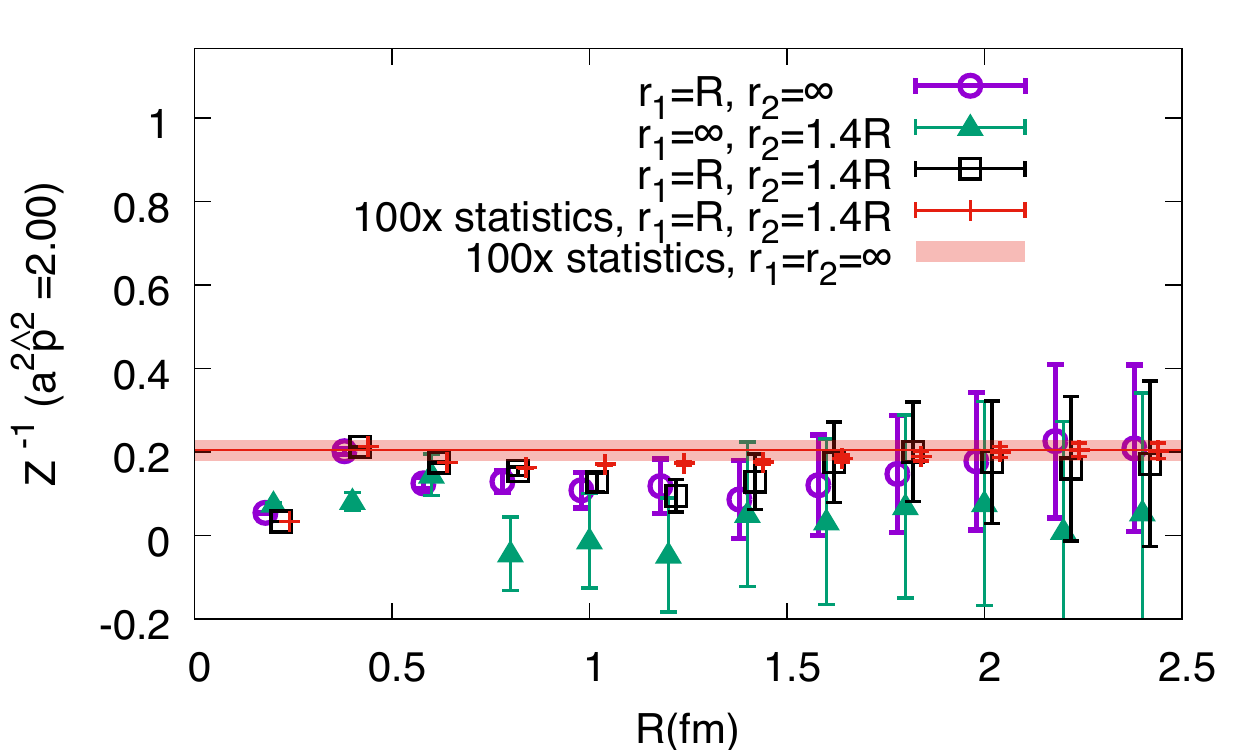}
   \includegraphics[scale=0.7]{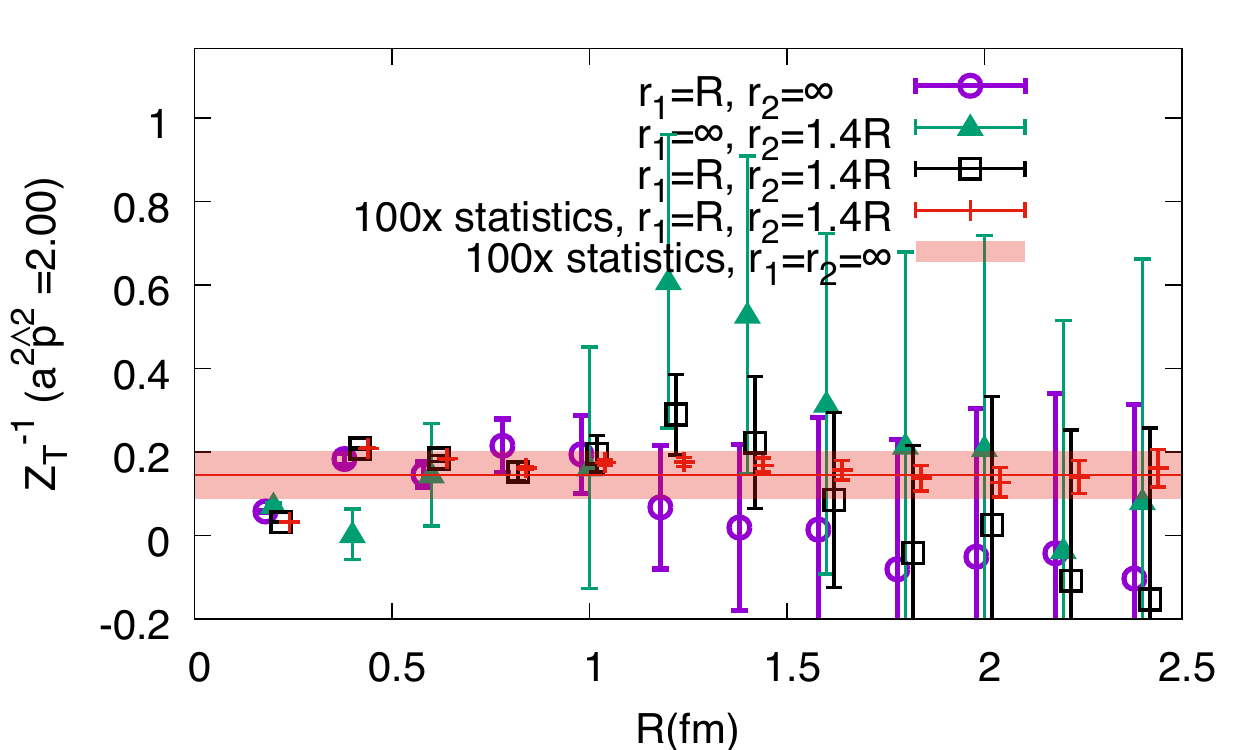}
 \caption{The cutoff $R$  dependence for $r_{1,2}$ of the renormalization constant $Z^{-1}(2\text{ GeV})$ and $Z_T^{-1}(2\text{ GeV})$ on the 24Q ensemble with $a^2p^2=2.00$. Calculation on 300 configurations with $r_1\ge0.7$~fm and $r_2\ge1.0$~fm are consistent with those using 70,834 configurations without any cutoff. The result is less sensitive to the cutoff $r_1$ than $r_2$; thus, most of the variance reduction comes from reducing $r_1$, while reducing $r_2$ is also useful. The green/black/red data are shifted horizontally to enhance legibility.}\label{fig:Z_quench2}
\end{figure}

\begin{figure}[ht]
  \includegraphics[scale=0.7]{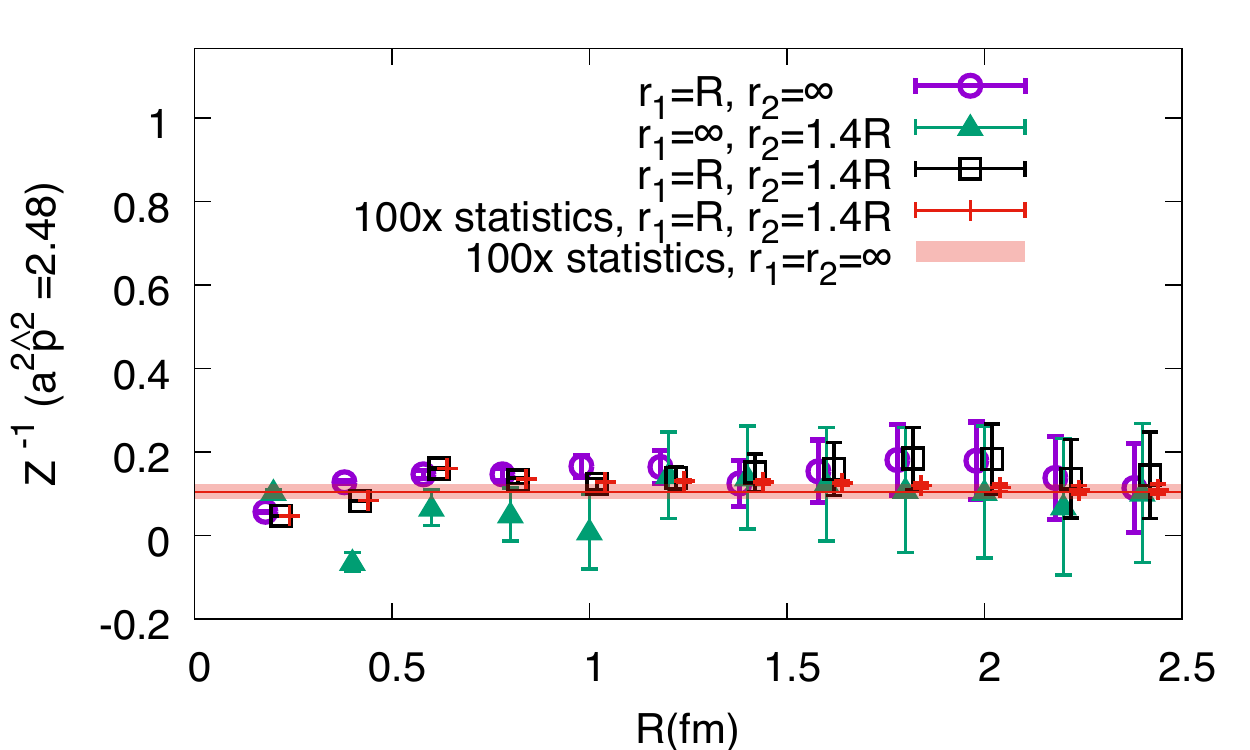}
   \includegraphics[scale=0.7]{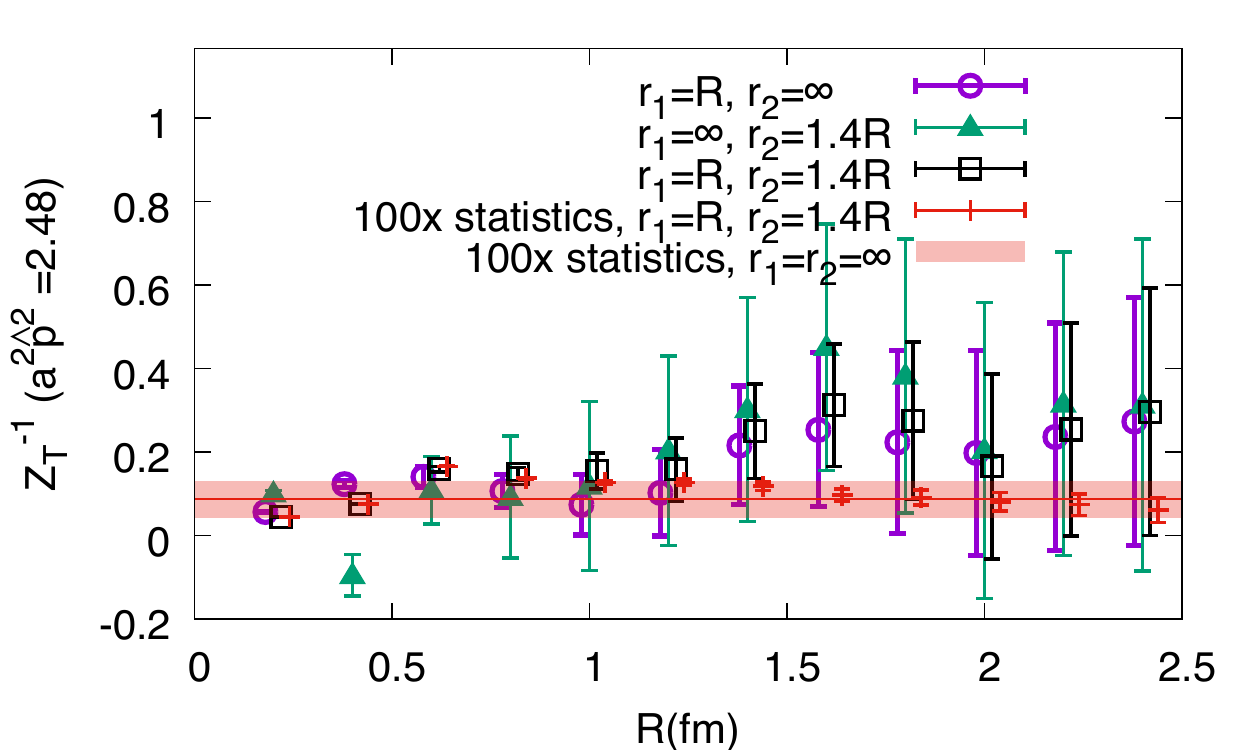}
 \caption{The cutoff $R$ dependence of the renormalization constant $Z^{-1}(2\text{ GeV})$ and $Z_T^{-1}(2\text{ GeV})$ on the 24Q ensemble with $a^2p^2=2.48$.}\label{fig:Z_quench3}
\end{figure}

\begin{figure}[ht]
  \includegraphics[scale=0.7]{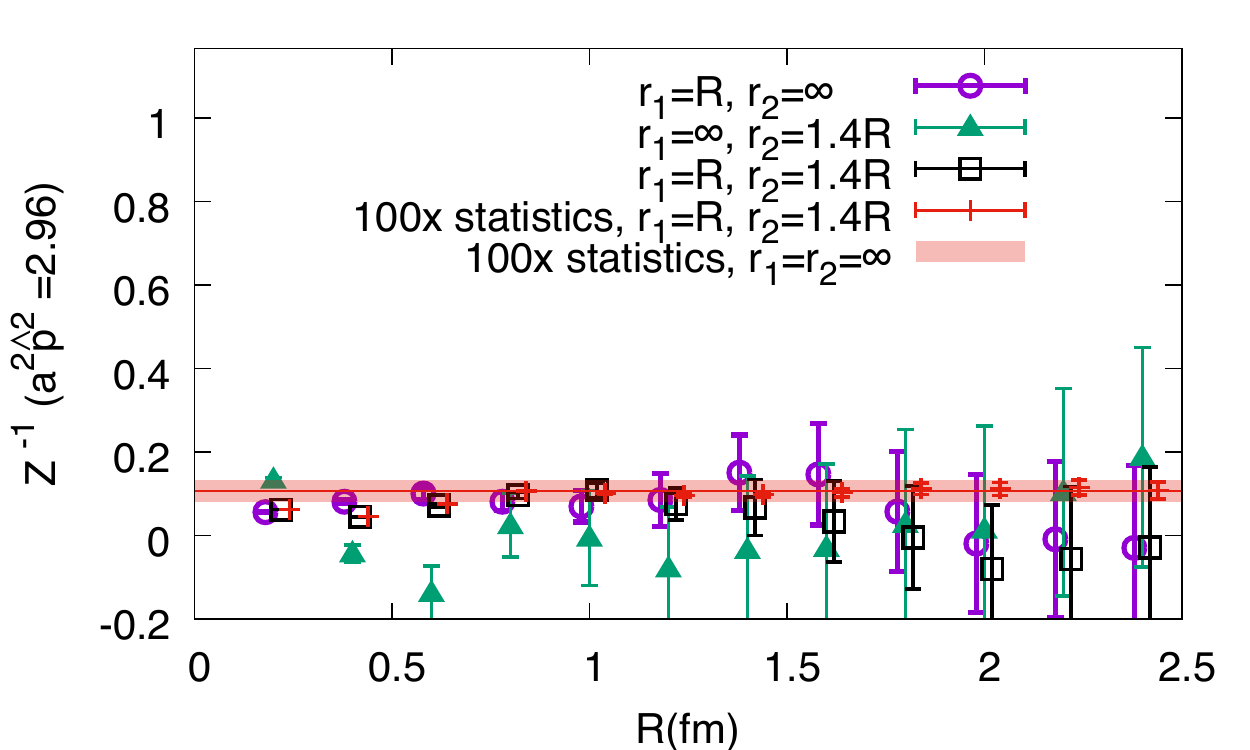}
   \includegraphics[scale=0.7]{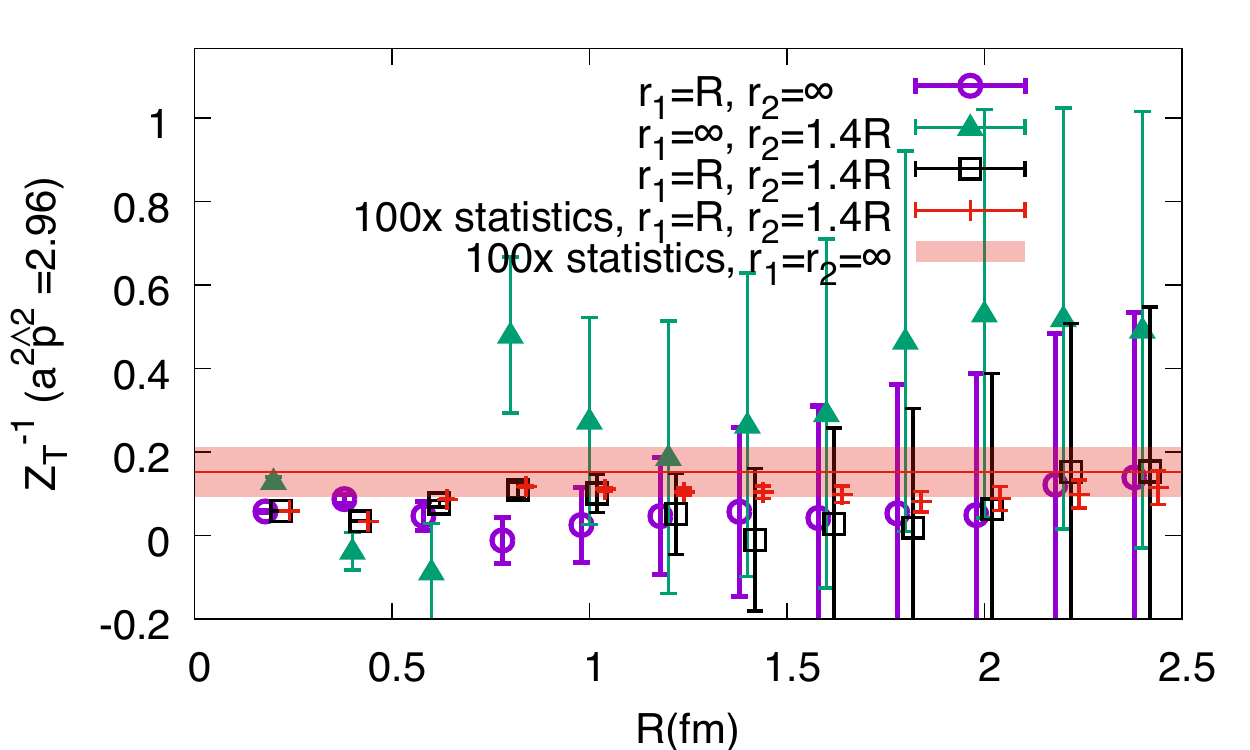}
 \caption{The cutoff $R$ dependence of the renormalization constant $Z^{-1}(2\text{ GeV})$ and  $Z_T^{-1}(2\text{ GeV})$ on the 24Q ensemble with $a^2p^2=2.96$.}\label{fig:Z_quench4}
\end{figure}

\subsection{Quenched ensemble 24Q}

We calculated $Z(1\text{ HYP})$ without CDER on decorrelated 70,834 configurations of a quenched Wilson gauge ensemble ``24Q'' with $a=0.098$~fm and $L^3\times V=24^3\times 64$, and compared them with those on 708 of the 70,834 configurations (pick 1 per 100 configuration numbers) with CDER. The CDER results with $r_1\ge0.8$~fm and $r_2\ge1.1$~fm agree with the CDER-free results for all $a^2\hat{p}^2$. Fig~\ref{fig:Z_quench2}, \ref{fig:Z_quench3} and \ref{fig:Z_quench4} show the $Z^{-1}$ and $Z_T^{-1}$ results with $a^2\hat{p}^2$=2.00, 2.48 and 3.00 respectively. In those figures, the red bands show the results on 70,834 configurations without CDER, and the black boxes show the results with $r_1=0.7\times r_2=R$ agree with the red bands for all the $R$'s not smaller than 0.7~fm. Result with the cutoff on either $r_1$ or $r_2$ set to $\infty$ (the green triangles and purple dots) are also shown in the figures, and it is obvious from the most left data points that the cutoff effects on $r_2$ are as strong as those on $r_1$ when $r_1=0.7\times r_2$. Thus setting the $r_{1,2}$ with this relation can be a proper choice to simplify the parameter tuning. The results also demonstrate that cutoffs on either $r_1$ or $r_2$ also reduce the statistical uncertainties of $Z^{-1}$. As shown in Fig.~\ref{fig:Z_quench2}-\ref{fig:Z_quench4}, the full statistics CDER results (red crosses) actually saturate at $R>$0.8 fm or so, and are consistent with both the full statistics non-CDER results and the 1\% statistics CDER results as expected.


\subsection{2 flavor ensembles 24C/12C}

\begin{figure}[ht]
  \includegraphics[scale=0.7]{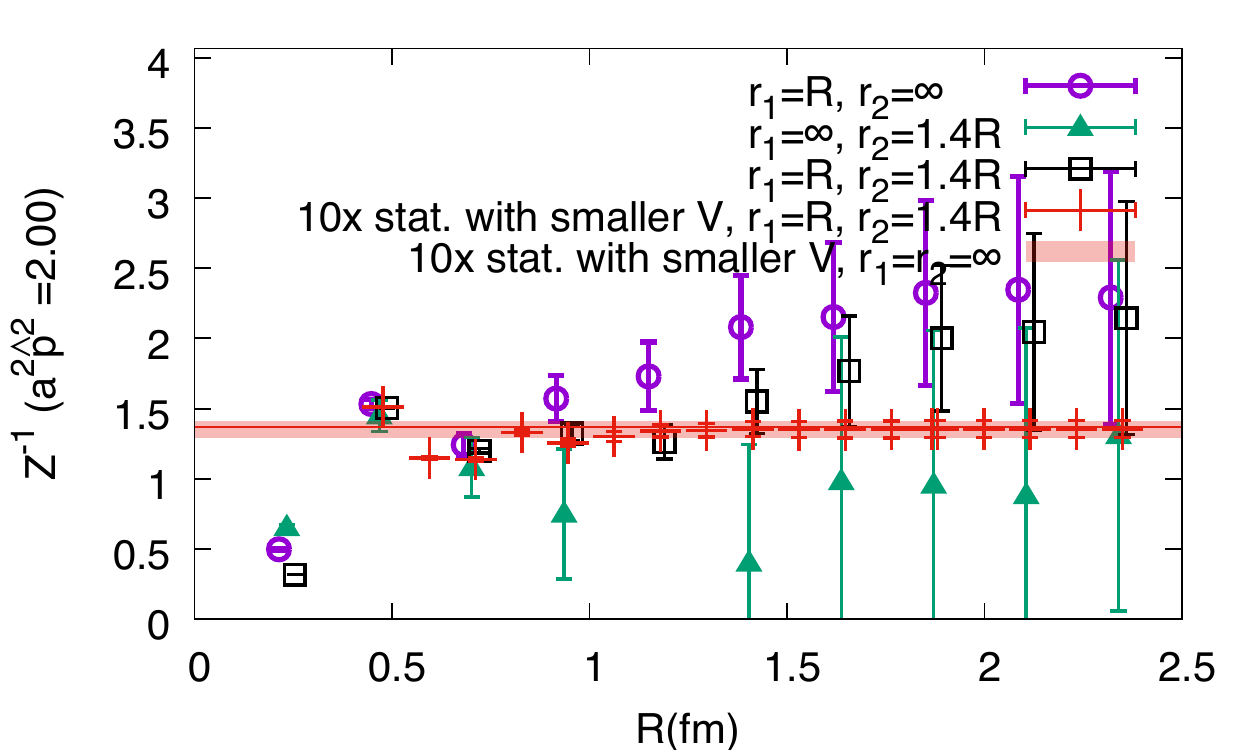}
   \includegraphics[scale=0.7]{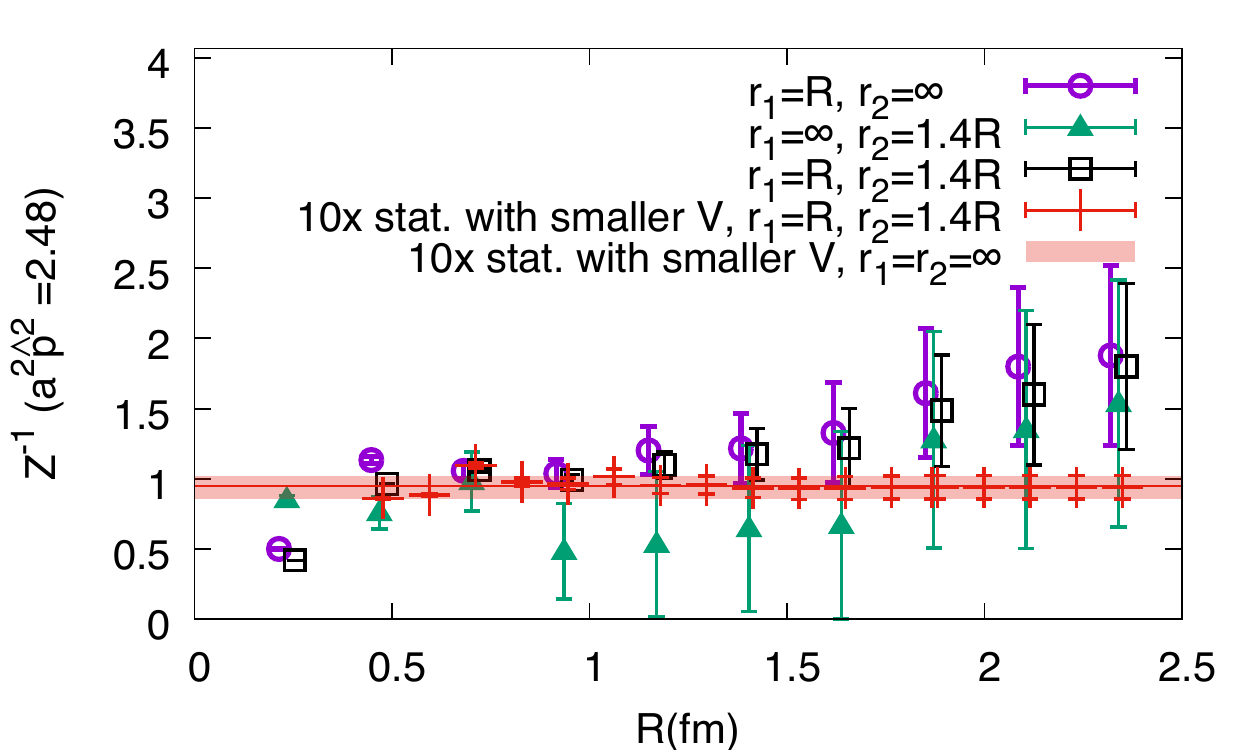}
 \caption{The cutoff $R$ dependence of the renormalization constant $Z^{-1}(2\text{ GeV})$ on the 24C/12C ensembles with $a^2p^2=2.00$ and 2.48.}\label{fig:Z_mit2}
\end{figure}

We also studied the dynamical case: we calculate $Z^{-1}(1\text{ HYP})$ with CDER on 2,123 configurations of the 2-flavor clover fermion L\"uscher-Weisz gauge ensemble ``24C'' with lattice spacing $a=0.117$~fm, $m_{\pi}=450$~MeV and $L^3\times V=24^3\times 64$~\cite{Orginos:2015aya}. For comparison, we repeat the calculation of $Z^{-1}$ on 21,166 configurations on the ``12C'' ensemble (with the same lattice setup as 24C except a smaller volume $12^3\times 24$) without CDER. Fig.~\ref{fig:Z_mit2} shows similar $R$-dependence plots for the dynamical case with 24C and 12C lattices ($a=0.117$~fm, $m_{\pi}=450$~MeV,  $L^3\times V$ equal to $24^3\times 64$ and $12^3\times 24$ respectively).  The red bands show the results on 21,166 configurations without any cutoff, and the data points show the CDER-results. They are all consistent for all the $R$'s not smaller than 0.9~fm. The uncertainty of the full statistics CDER results are not much smaller than the non-CDER ones since the volume is too small to make the CDER efficient.

For the cutoffs on the radii $r_1$ and $r_2$, they should correspond to the respective correlation lengths between the relevant operators. $r_1$ is between the gauge field and the EMT operator. Taking the vector meson $\omega(780)$ as an estimate, the correlation length $3/m_{\omega}\sim$ 0.76 fm (at 3 times the Compton wavelength, the Yukawa potential has fallen by 95\%) is close to 0.9 fm that we take for $r_1$. 
 On the other hand, the gluon has a "dynamical mass" $m_g\sim$ 550 MeV in the small momentum region~\cite{Oliveira:2010xc,Cucchieri:2011ig}. This gives an estimate of the correlation length of $3/m_g\sim$ 1.2 fm which is close to the 1.3 fm cutoff used for $r_2$.

\begin{figure}[ht]
\includegraphics[scale=0.7]{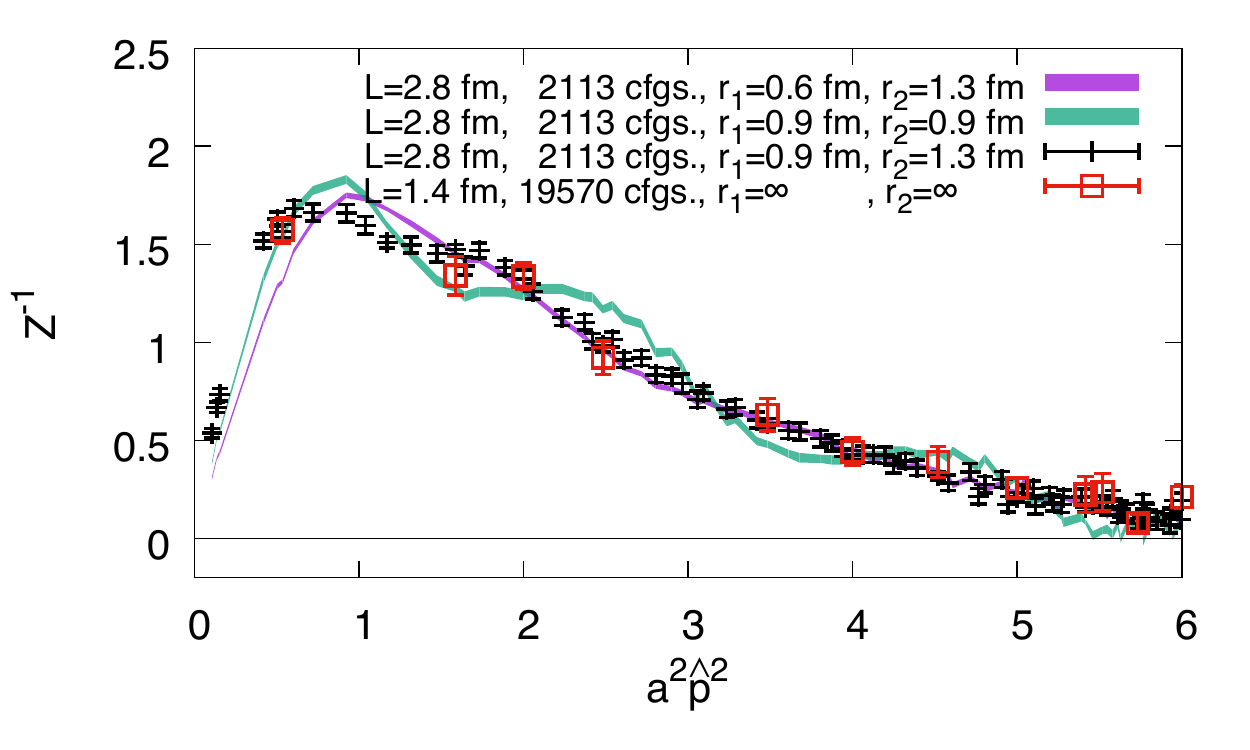}
\caption{The $\MSbar$ renormalization constant $Z^{-1}(2\text{ GeV})$ on the 24C ensemble as a function of $a^2\hat{p}^2$, with different cutoffs on the gluon field-operator correlation ($r_1$) and propagator ($r_2$). A high-statistics calculation without cutoff on a lattice with smaller volume but the same paramters is also presented (red boxes) for comparison.}\label{fig:Z_2f}
\end{figure}

As in Fig.~\ref{fig:Z_2f}, we should choose $r_1 \ge 0.9$ fm and $r_2 \ge 1.3$ fm on 24C (black crosses) to get the consistent results with those on 12C without CDER (the red boxes). If we fit the CDER-result of $Z^{-1}$ on 24C with a polynomial form including $a^{2n}\hat{p}^{2n}$ ($n\le$2) terms in the range $a^{2}\hat{p}^{2}\in$[1.5, 5], the result is 2.63(5) with $\chi^2$/d.o.f.=0.80. Fig.~\ref{fig:Z_2f} also shows $Z^{-1}(1\text{ HYP})$ with either smaller $r_1$ (the purple band) or $r_2$ (the green band); one can see that although the statistical uncertainties are smaller, there are distinct systematic bias in the form of oscillations in $a^2\hat{p}^2$ and the $\chi^2$/d.o.f. with similar fitting setup will be 6.1 and 28.9 respectively. Thus the $\chi^2$/d.o.f. can provide a consistent criteria on the systematic uncertainties introduced by CDER, especially in the case (likes 48I) we cannot resolve any signal without CDER.

\subsection{2+1 flavor ensembles 48I}

\begin{figure}[ht]
\includegraphics[scale=0.7]{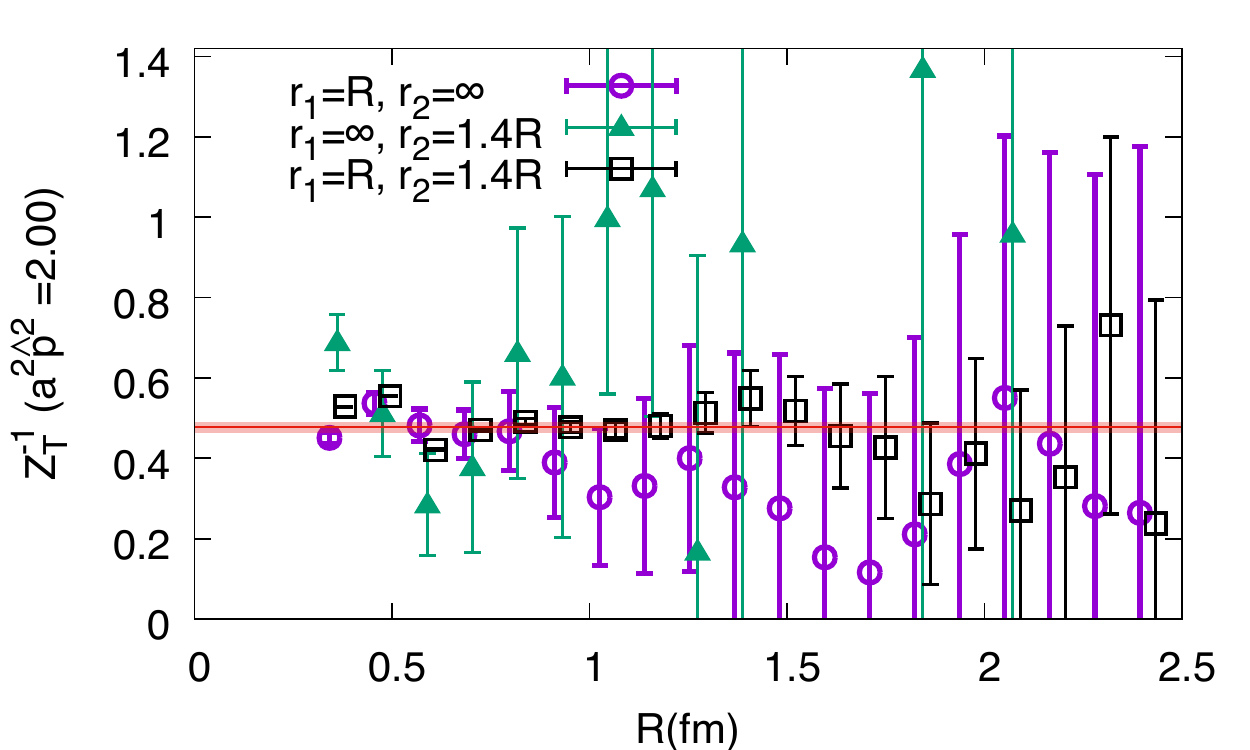}
\includegraphics[scale=0.7]{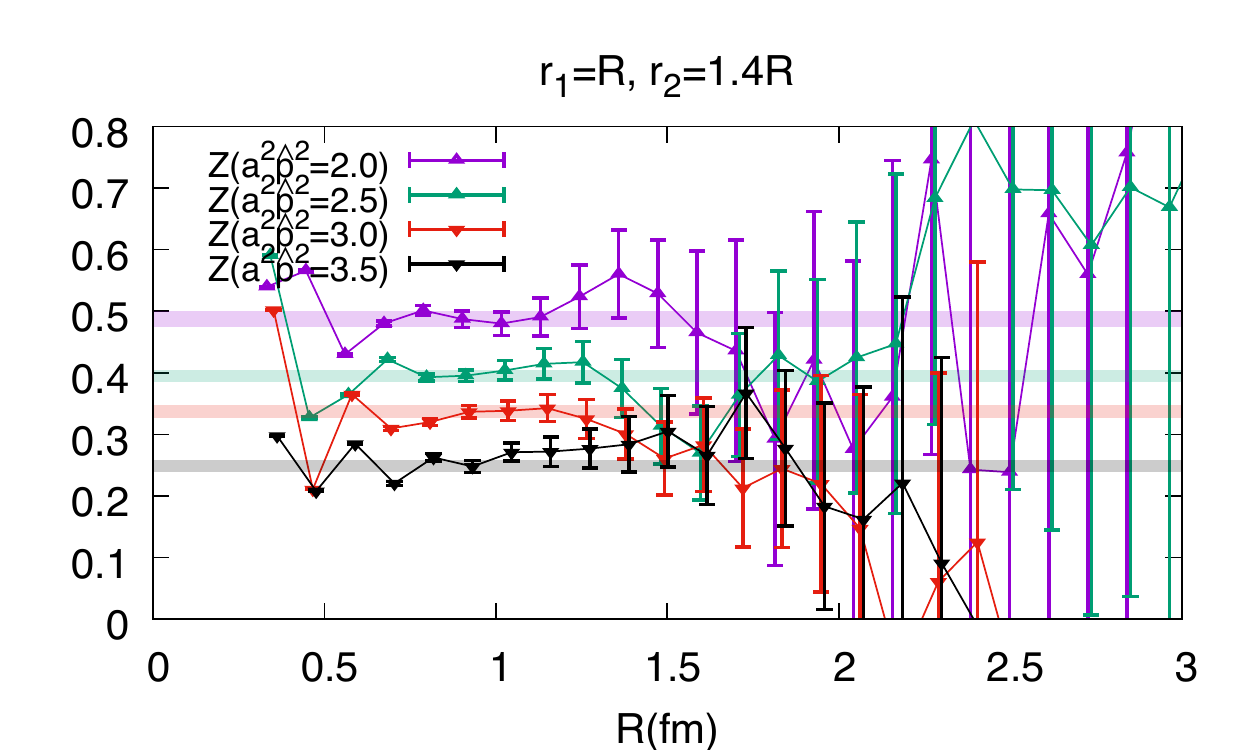}
\includegraphics[scale=0.7]{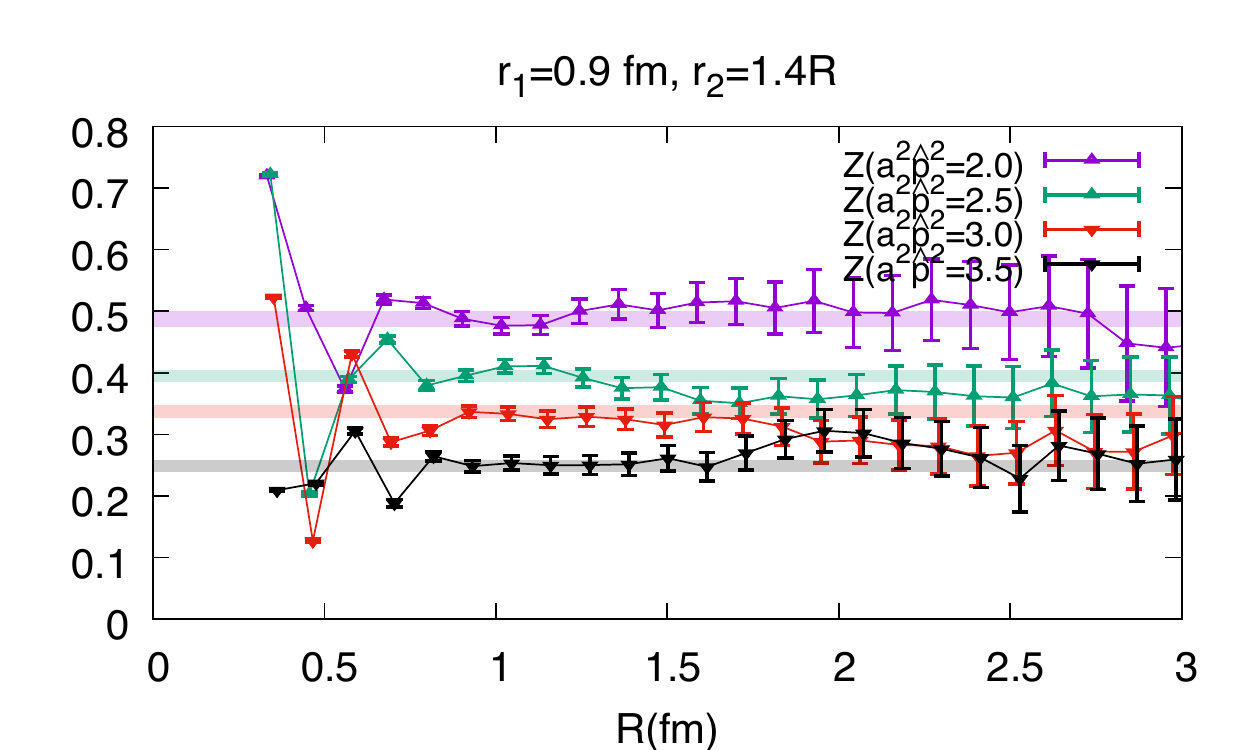}
\caption{The cutoff $R$ dependence of the renormalization constant $Z_T^{-1}(2\text{ GeV})$ on the 48I ensemble. }\label{fig:48I_cut}
\end{figure}

{Before the end of this section, a few $R$-dependence tests on 48I, the ensemble we will use for the final result, are provided in Fig.~\ref{fig:48I_cut}.} 

In the {upper} panel of Fig.~\ref{fig:48I_cut},  the $Z_T^{-1}(2\text{ GeV})$ case with $a^2p^2$=2.00 is presented using the similar style as the previous plots in this section, while the bands are based on the results with the CDER cutoffs $r_1$=$0.7\times r_2$=0.9 fm. It is obvious that the cutoffs on $r_{1,2}$ are necessary as the errors with either $r_1$ or $r_2$ cutoff only are very large. The {central} panel of Fig.~\ref{fig:48I_cut} shows the cutoff $R$ dependence with $r_1=0.7\times r_2=R$ at $a^2p^2$=2.00, 2.48, 2.97,  and 3.48. All the data points with $R>$ 0.9 fm are consistent with the band based on the data point at $R=0.9$. {In the lower panel of Fig. ~\ref{fig:48I_cut}, the cutoff $R=0.7\times r_2$ dependences at different $a^2\hat{p}^2$ are presented with  fixed $r_1=$0.9 fm. Thus the uncertainty with larger $r_2$ are smaller and then consistency is more obvious.
As an estimate of the systematic uncertainty due to the choice of $r_2$, we take the 2\% fluctuation of the gluon
propagator at $r_2 = 1.3$ fm as the systematic error in our final prediction.}
}
\section{Renormalized $\xg$ on 48I}\label{sec:result}

\begin{figure}[ht]
\includegraphics[scale=0.7]{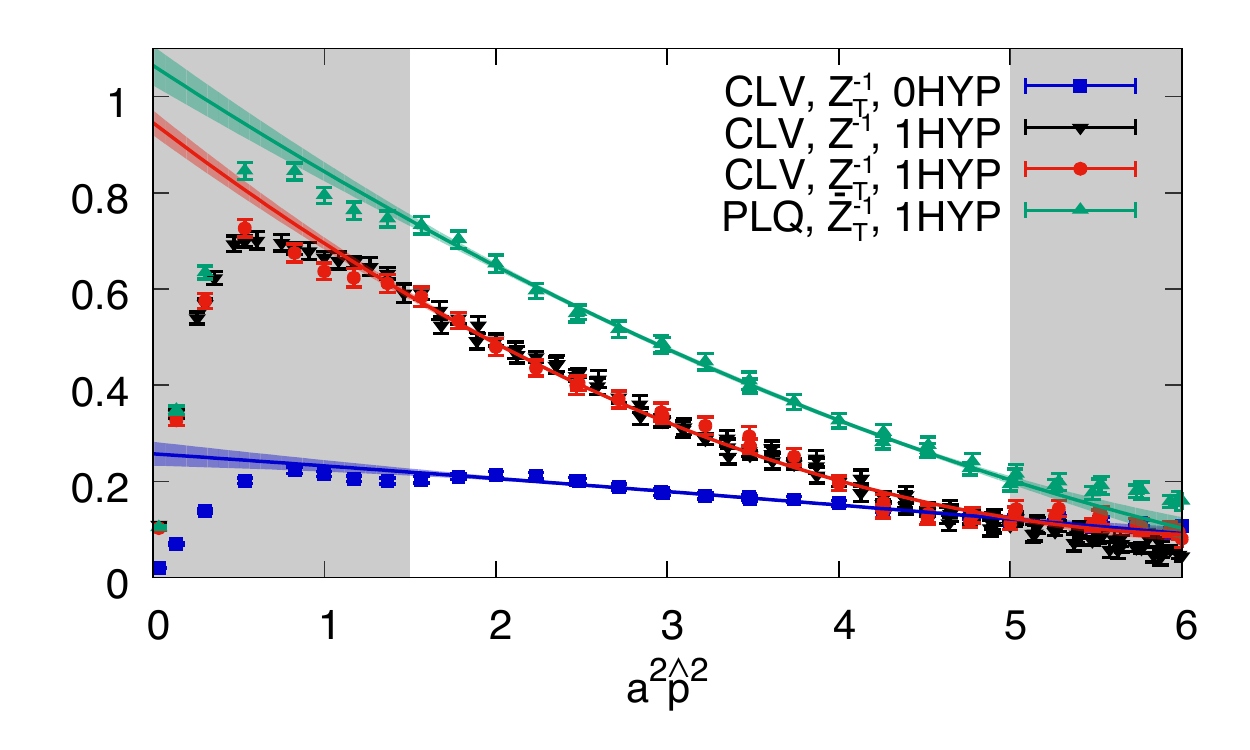}
\caption{The $\MSbar$ at 2~GeV renormalization constants as functions of $a^2\hat{p}^2$, for the gauge EMT operators. The red dots and blue boxes show the $Z_T^{-1}$ with and without HYP smearing using the clover definition (CLV), and the green triangles show the HYP smeared case using the plaquette definition (PLQ). The result of $Z^{-1}$ with HYP smearing and the clover definition (purple triangles) are also plotted for the comparison.}\label{fig:Z_g_final}
\end{figure}

  Given the success of CDER in resolving clean signal of $Z^{-1}_T$, it is nevertheless important to confirm that the renormalized $\xg$ is independent of the lattice definition of $\ocT_{g}$ or whether the HYP smearing is applied, up to ${\cal O}(a^2)$ corrections. Fig.~\ref{fig:Z_g_final} gives the CDER-results on the 48I ensemble as the functions of $a^2\hat{p}^2$. The red dots and blue boxes show $Z^{-1}_{T}$ with and without HYP smearing, respectively, using the clover definition in Eq.~(\ref{eq:emt_clover}); the green triangles show the HYP-smeared case using the plaquette definition in Eq.~(\ref{eq:emt_plq}), $\bar{Z}^{-1}_{T}$. The $a^2\hat{p}^2$ dependence and the $a^2\hat{p}^2\to 0$ limit of the renormalization constants are different between the different definitions, while the presumed rotation symmetry breaking between $Z^{-1}$ (black triangles) and $Z^{-1}_{T}$ are consistent with zero within the uncertainties. With the functional form $Z_T^{-1}(a^2\hat{p}^2)=Z_T^{-1}(0)+C_1a^2\hat{p}^2+C_2a^4\hat{p}^4$, we fit the range $a^2\hat{p}^2\in[1.5, 5]$ (the lighter area in Fig.~~\ref{fig:Z_g_final}) and obtain $Z^{-1}_{T}(0\text{ HYP})=0.257(25)(5)$, $Z^{-1}_{T}(1\text{ HYP})=0.946(26)(19)$ and $\bar{Z}^{-1}_{T}(1\text{ HYP})=1.05(35)(21)$ where the second error is an estimate of the systematic uncertainty from the 2\% truncation error of the gluon propagator at $r_2\sim$1.3 fm. The $\chi/d.o.f.$ for all the cases are smaller than 1.

\begin{figure}[ht]
\includegraphics[scale=0.7]{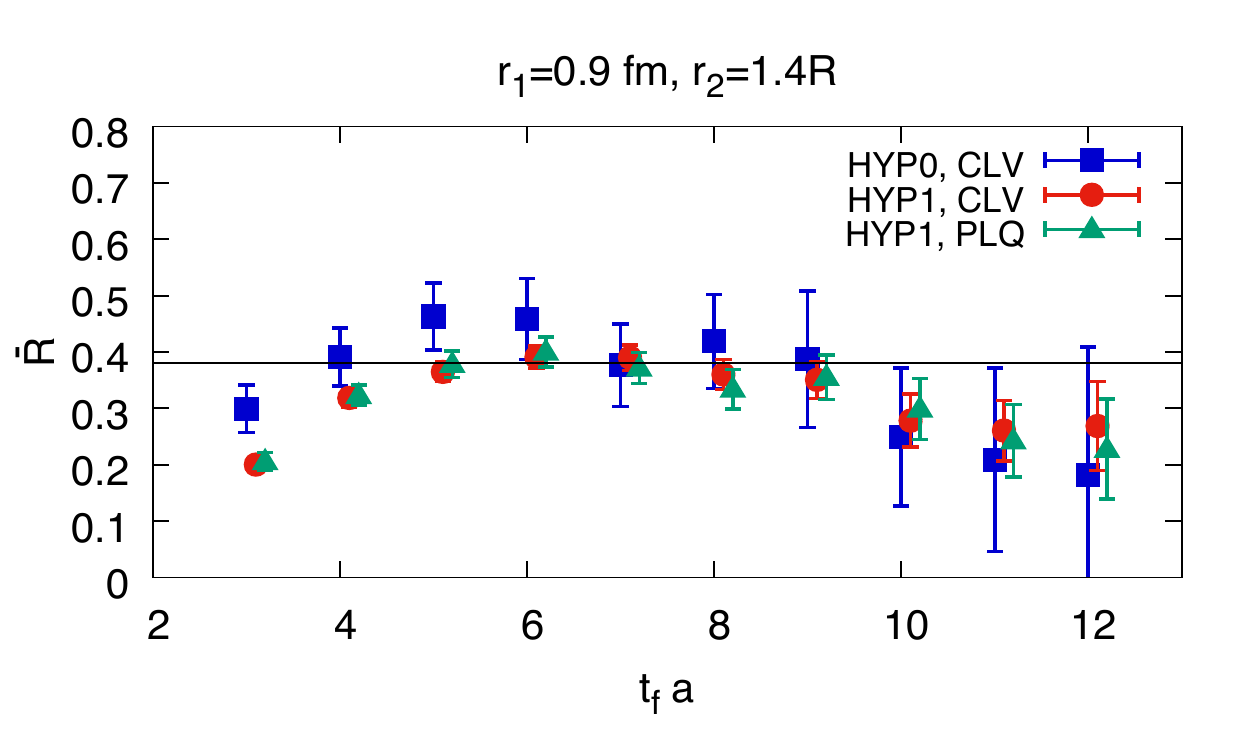}
\caption{The renormalized $\bar{R}(t_f)$ with and without the HYP smearing (the red dots and blue boxes respectively) using the clover definition, and also the HYP-smeared case with the plaquette definition (the green triangles). The HYP-smeared data are shifted horizontally to enhance the legibility and a black line at 0.38 is placed on the figure to guide the eyes. All results agree with each other within $2\sigma$ for $t_f\ge 4$.}\label{fig:renorm_me}
\end{figure}

To determine the bare $\xg$, the following ratio is calculated in the rest frame of the nucleon on 81 configurations of the 48I ensemble with partially quenched valence overlap fermion for the pion mass $m_{\pi}\in[135, 372]$~MeV,
\begin{equation}
R(t_f,t) =
  \frac{4\langle 0|\Gamma^e\int d^3y\,\chi(\vec{y},t_f) \ocT_{g,44}(t)\bar{\chi}(\vec{0},0)|0 \rangle}
			 {3M_N\langle 0|\Gamma^e\int d^3y\,\chi(\vec{y},t_f)\bar{\chi}(\vec{0},0)|0 \rangle},
\end{equation}
where $\chi$ is the nucleon interpolation field, $\Gamma^e$ is the unpolarized projection operator of the proton and $M_N$ is the nucleon mass. When $t_f$ is large enough, the derivative of the $t$-summed ratio $R(t_f,t)$ becomes the glue momentum fraction in the nucleon, as applied in the recent high-accuracy nucleon matrix element calculation~\cite{Berkowitz:2017gql},
\begin{align}
\bar{R}(t_f) \equiv{}&\sum_{0<t<t_f}R(t_f,t)-\sum_{0<t<t_f-1}R(t_f-1,t)\nonumber\\
={}&\xg^\text{bare}+{\cal O}(e^{-\delta m\,t_f}),
\end{align}
up to the excited-state contamination at ${\cal O}(e^{-\delta m\,t_f})$. The calculation setup is the same as for our previous work on the glue spin~\cite{Yang:2016plb}: a $4\times 4 \times 4$ smeared grid source with low mode substitution (LMS)~\cite{Li:2010pw} is used for the nucleon two-point functions, and all the time slices are looped over to increase statistics. We followed the same strategy in Ref.~\cite{Liu:2017man} to apply CDER to the numerator of $R(t_f,t)$. With a cutoff around 1~fm 
which is enough as demonstrated in the the NPR cases studied here, the statistical uncertainties of $\bar{R}(t_f)$ can be reduced by an factor of $\sim$10. The systematic uncertainties in bare $\bar{R}(t_f)$ due to CDER will be investigated in the future following the strategy in Ref.~\cite{Liu:2017man}. The renormalized $\bar{R}(t_f)$ at $m_{\pi}=372$~MeV is shown in Fig.~\ref{fig:renorm_me} as a check with the best signals we have. The errors from $Z_T$ and the bare $\bar{R}(t_f)$ are combined in quadrature. As shown in that figure, even though the renormalization constants with or without HYP smearing differ by a factor of $\sim$3 as we saw in Fig.~\ref{fig:Z_g_final}, the renormalized $\bar{R}^R(t_f)\equiv Z_T\bar{R}(t_f)$ are consistent within $2\sigma$ for $t_f\ge4$.

\begin{figure}[ht]
\includegraphics[scale=0.7]{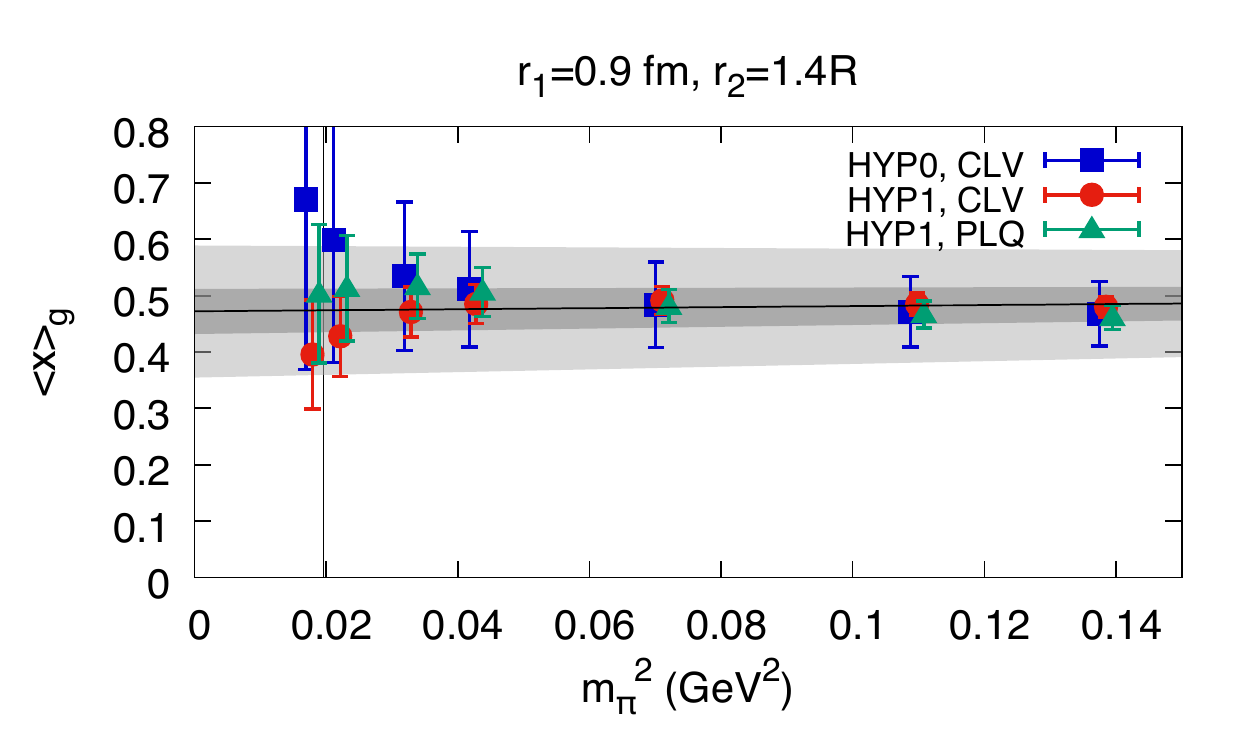}
\caption{The renormalized glue momentum fraction $\xg$ as a function of $m_{\pi}^2$. The HYP-smeared data are slightly shifted horizontally to enhance the legibility.  The results with different definitions are consistent with each other and the $m_{\pi}^2$ dependence is mild. The dark and light gray bands shows the statistical and total uncertainties respectively at a combined linear fit of the $m_{\pi}^2$ dependence.}\label{fig:x_g_final}
\end{figure}

 We fit $\bar{R}(t_f)$ to a constant in the range $t_f\ge7a$ to obtain $\xg$ and plot its $m_{\pi}^2$ dependence in Fig.~\ref{fig:x_g_final}. With a linear fit to $m_{\pi}^2$ for $m_{\pi}<400$~MeV on the 1-HYP smeared data with the clover definition, we obtain $\xg^{\overline{\textrm{MS}}}$(2~GeV) at the physical pion mass to be $0.47(4)(11)$. The variance of the values from three definitions, the uncertainties of the renormalization constants, and the mixing effect from the quark momentum fraction $\langle x \rangle_q$ (which is estimated by $1-\xg$ times the 1-loop mixing coefficient $0.1528$~\cite{Yang:2016xsb}) are combined in quadrature as the systematic uncertainty.  The prediction is consistent with the global fitting result CT14 \cite{Dulat:2015mca} 0.42(2) in $\overline{\textrm{MS}}$ at the same scale. The major systematic uncertainty is the mixing from the quark and can be eliminated with a similar non-perurabtive calculation with the quark external states. 
\\


\section{Summary}\label{sec:summary}

In summary, we have presented a systematic implementation of NPR for the glue momentum fraction $\xg$. We demonstrated that the CDER technique can provide an unbiased improvement on the lattice with the cutoffs $r_1\sim0.9$ fm and $r_2 \sim1.3$~fm, and that the renormalized $\xg$ is insensitive to the lattice definition of the gauge EMT or HYP smearing within uncertainties.

Our calculation also shows that HYP smearing can make the $a^2\hat{p}^2$ dependence of the renormalization constant much stronger than the case without HYP smearing, even though the $a^2\hat{p}^2$-extrapolated value can be closer to one. 
The cases with more steps of HYP smearing is shown in the {appendix}.

\hspace{1cm}
\section*{Acknowledgments}

We thank W. Detmold, L. Jin and P. Sun for useful discussions, 
and the RBC and UKQCD collaborations for providing
us their DWF gauge configurations. HL and YY are
supported by the US National Science Foundation under
grant PHY 1653405 ``CAREER: Constraining Parton
Distribution Functions for New-Physics Searches".
This work is partially supported by DOE grant DE-SC0013065 and DOE TMD topical collaboration. This research
used resources of the Oak Ridge Leadership Computing
Facility at the Oak Ridge National Laboratory,
which is supported by the Office of Science of the U.S.
Department of Energy under Contract No. DE-AC05-00OR22725. This work used Stampede time under the
Extreme Science and Engineering Discovery Environment
(XSEDE), which is supported by National Science Foundation
grant number ACI-1053575. We also thank National
Energy Research Scientific Computing Center (NERSC)
for providing HPC resources that have contributed to the
research results reported within this paper. We acknowledge
the facilities of the USQCD Collaboration used for
this research in part, which are funded by the Office of
Science of the U.S. Department of Energy. 


\begin{widetext}
{
\section*{Appendix: The discretization error with more steps of HYP smearing}\label{sec:appendix}

In this section, we repeat the NPR and matrix elements calculation on 48I, but with 2 and 5 steps of HYP smearing. 

As shown in the left panel of Fig.~\ref{fig:2hyp}, $Z^{-1}_T$ becomes increasing non-linear on $a^2\hat{p}^2$ when more HYP smearing steps are applied on the gauge EMT. Without HYP smearing, the $a^2\hat{p}^2$ dependence of $Z^{-1}_T$ can be well described by a linear term and the coefficient of the next order $a^4\hat{p}^4$ term is consistent with zero. With more HYP smearing steps, the coefficients of the $a^2\hat{p}^2$ and $a^4\hat{p}^4$ terms increase significantly. Since all momenta $p$ on the external legs of the gauge EMT will be integrated in the hadron matrix element, $a^{2n}\hat{p}^{2n}$ corrections will result in ${\cal O}(a^{2n})$ discretization errors at finite lattice spacing. From the renormalized $\bar{R}(t_f)$ in the right panel of Fig.~\ref{fig:2hyp}, the results with 2 steps of HYP smearing still agree with the results with 1 step of HYP smearing; but if we jump to the 5-step HYP smearing used by some previous studies, the $a^{2n}\hat{p}^{2n}$ corrections will be much larger and the renormalized result will be have large systematic uncertainties from determining $Z^{-1}_T$ (green triangles and blue boxes).

In the 5HYP case, with the same range $a^2\hat{p}^2\in$ [1.5, 5] and the polynomial form up to $a^4\hat{p}^4$ term,  $Z^{-1}_T$(5HYP)=0.663(35) is obtained with $\chi^2=0.8$ (the default fit). If the $a^6\hat{p}^6$ term is added and the range is switched to $a^2\hat{p}^2\in$ [1, 4],  $Z^{-1}_T$(5HYP) will jump to 1.11(11) with $\chi^2=0.4$ (the tuned fit). The data of  $Z^{-1}_T$(5HYP) (the green triangles) with the band from the default fit (the green band) and tuned fit (the blue band) are plotted in the left panel of Fig.~\ref{fig:2hyp}, and the renormalized $\bar{R}(t_f)$ with both fits of $Z^{-1}_T$ are shown in the right panel. The errors from $Z_T$ and the bare $\bar{R}(t_f)$ are combined in quadrature. It is obvious that the renormalized $\bar{R}(t_f)$ with 5-step HYP smearing  (green triangles) based on the default fit of $Z^{-1}_T$ is much higher than those with 1,2 steps of HYP smearing. Even though the consistency can be improved if the tuned fit of $Z^{-1}_T$  is applied (the blue boxes), the systematic uncertainties from the fit of $Z^{-1}_T$ will make the final uncertainties in the 5-step HYP smearing case larger than the cases with fewer steps of HYP smearing.

\begin{figure}[ht]
  \includegraphics[scale=0.7]{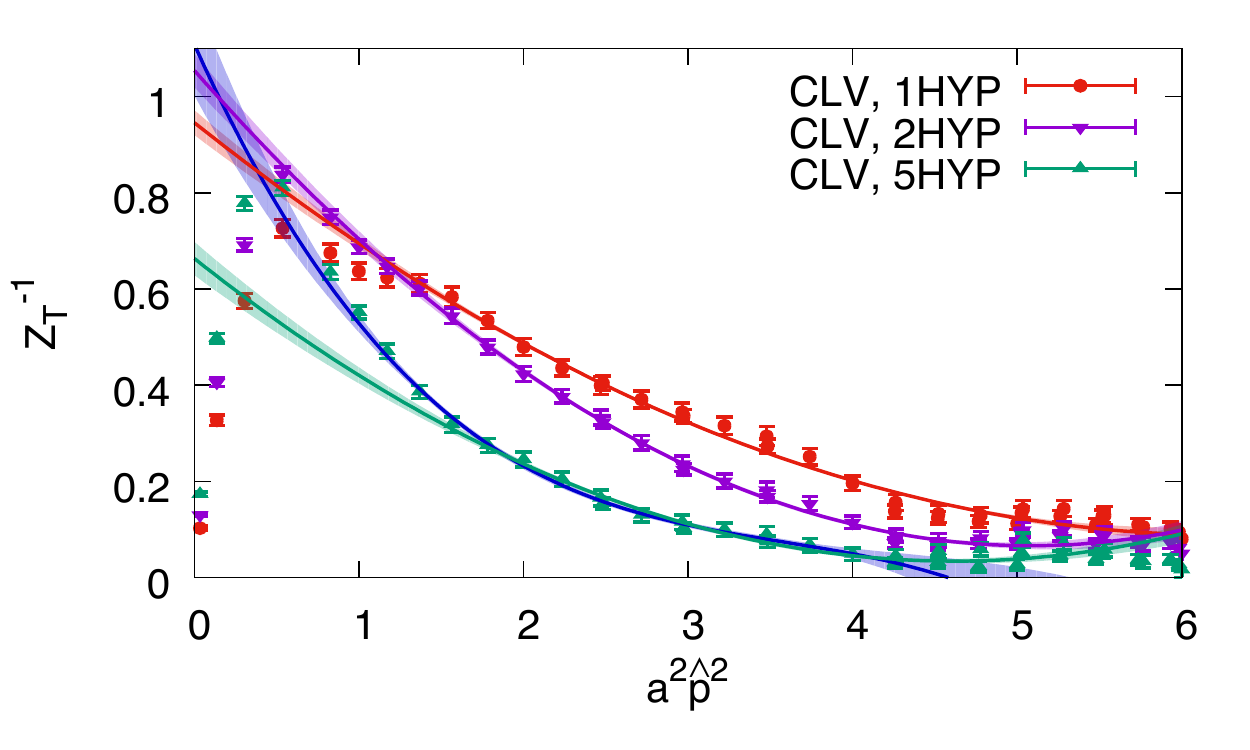}
   \includegraphics[scale=0.7]{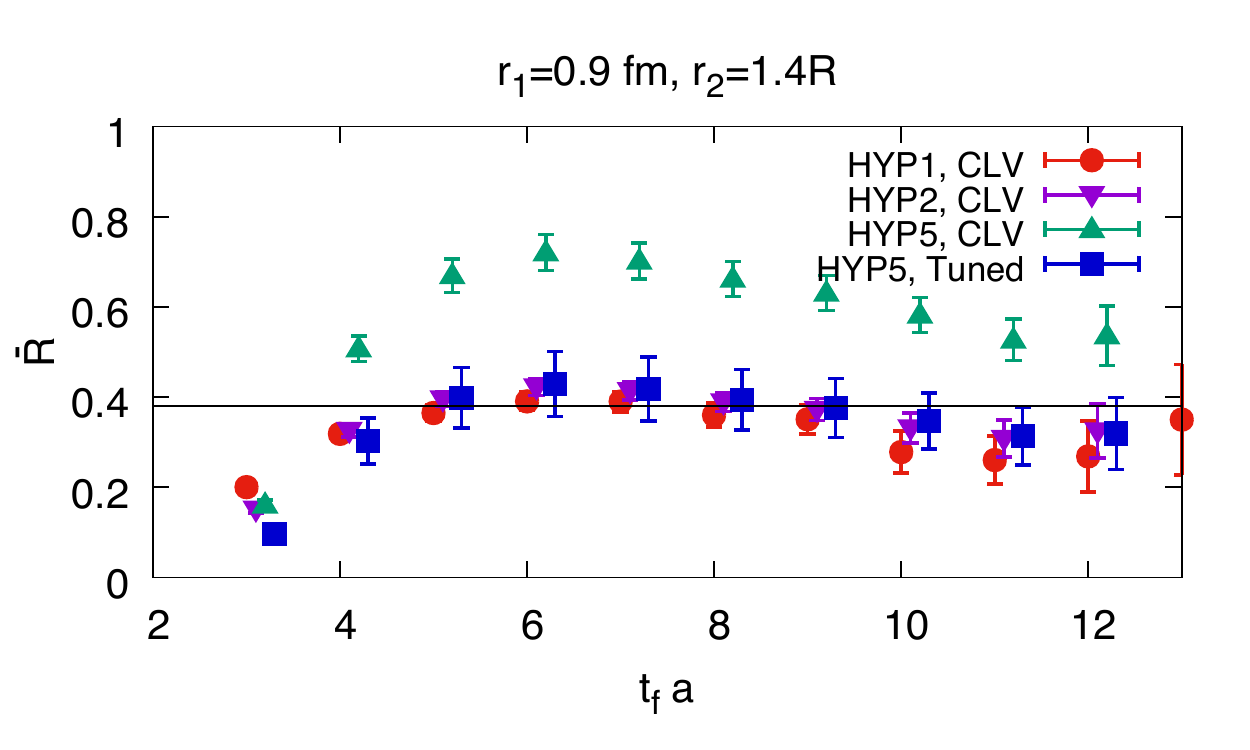}
 \caption{The $\overline{\textrm{MS}}$ 2GeV renormalization constants $Z^{-1}_T$ and renormalized $\bar{R}(t_f)$ with 1,2,5 steps of HYP smearing are shown as the red dots, purple reversed triangles and green triangles, respectively. Both the blue and green bands are the fit of the 5-HYP data, with the regions $a^2\hat{p}^2\in$ [1, 4] and [1.5,5] respectively. The renormalized $\bar{R}(t_f)$ in the 2HYP case still consistent with the 1HYP case even though the $a^2\hat{p}^2$ dependence of $Z^{-1}_T$ are quite different for $t_f\ge$3, but the 5HYP case will be very sensitive to the fit of $Z^{-1}_T$ and then has a large systematic uncertainties.}\label{fig:2hyp}
\end{figure}
}
\end{widetext}

\bibliographystyle{apsrev4-1}
\bibliography{reference.bib}

\end{document}